\documentclass{article}

\usepackage{graphicx}
\newcommand{\bra}[2]{\ensuremath{\langle#1|_{#2}}}
\newcommand{\ket}[2]{\ensuremath{|#1\rangle_{#2}}}

\title{\textbf{A defense of Hellwig-Kraus reductions}}  
\author{Richard A. Mould\\Department of Physics, State University of New York,\\Stony
Brook, New York, 11794-3800, USA}  
\date{}    

\begin{document}

\maketitle

\begin{abstract}

Aharonov and Albert analyze a thought experiment which they believe shows that
quantum mechanical state reductions occur along temporal hypersurfaces in Minkowski
space. They conclude that the covariant state reduction theory of Hellwig and Kraus
does not apply.  In Part I of this paper we disagree with this interpretation of
the A-A experiment, and show the adequacy of the H-K theory.  In Part II we examine the belief that H-K
reductions produce self contradicting causal loops, and/or give rise to absurd boundary conditions.  These
objections to the theory are shown to be unfounded.


\end{abstract}

\section{Introduction}

	When the position of a quantum mechanical particle is measured, its state undergoes
a collapse that is instantaneous and universal.  If that were not so, there would be
a finite probability of finding the particle in two different places at once.  The
collapse must furthermore be such that the particle cannot be simultaneously found
at two different places relative to any Lorentz observer.   It follows that state
reduction for a position measurement must be effective throughout the entire
space-like region surrounding the measurement.  This illustrates the
\emph{Hellwig-Kraus} reduction thesis which claims more generally that \emph{any}
measurement reduces the state in all of the  surrounding space-like  region \cite{HK}.
Accordingly, state reduction is felt all along the surface of the backward time cone, and at all events
forward of that surface. 

Several authors believe that there exist special measurements that do not result in
a Hellwig-Kraus type of reduction \cite{AA},\cite{CH},\cite{dE}.  The
reductions associated with these measurements are said to occur across hypersurfaces
of time, in violation of the covariant H-K reduction scheme.  The prototype of this
kind of measurement is given in a thought experiment proposed by Aharonov and Albert
in Ref.\ 2.    

\part{}

\section{The Aharonov \& Albert experment}

Two spatially separated 1/2 spin particles are initially prepared in the state

\begin{equation}
\ket{0,0}{} \ =\  \ket{J_z=0,J^2=0}{}\ =\ \frac{1}
{\sqrt{2}}\{\ket{\uparrow}{1}\ket{\downarrow}{2}-\ket{\downarrow}
{1}\ket{\uparrow}{2}\}
\end{equation}

The first particle is made to interact with a detector at point $x_1$, and the second
particle is made to interact with another detector at point $x_2 \not= x_1$, where the
interactions are simultaneous in some Lorentz Frame.  Prior to these interactions,
the detectors are brought together for the purpose of correlating their internal
variables in a specific way.  After the interactions, Aharonov \& Albert show that
neither one of the detectors has measured a single spin component of either particle
by itself.  Instead, the detectors have together recorded the fact that the
particles continue to be in the state \ket{0,0}{}.  On this basis, A\&A claim to
have designed ``\ldots a system of purely local experiments which measures a
nonlocal property of the system at a well defined time\ldots''(Ref.\ 2, p.363). 
That is true so long as ``measures'' refers only to a measurement \emph{interaction}, and
does not include the \emph{observation} that the author believes is necessary to produce the state
reduction associated with measurement.  A\&A do not make this distinction.  They focus entirely on the
interaction as though it is sufficient to bring about a state reduction.  They conclude that
this particular  reduction takes place across a (flat) hypersurface of time, which means
that the covariant (cone-shaped) reduction of Hellwig and Kraus cannot apply.

The author believes that a proper analysis of any state reduction must include an
account of all detector observations as well as their interactions.  Any observation is intrinsically  local, and is
said by von Neumann to trigger a measurement process that he calls Process I (see Sect. 5).  It is claimed here that
any such measurement will sharply locate the vertex of a Hellwig-Kraus (cone shaped) reduction at the observation
site; and therefore, that the A\&A experiment does not result in (flat) temporal hypersurface reductions.

\section{Nonlocal nondemolition}

	Consider the case shown in Fig.\ 1 in which one of the detector-particle interactions (event 1) takes place
before the other (event 2) in some Lorentz frame.  The world lines of particles
\#1 and \#2 are parallel to each other and to the \mbox{$t$-axis}, where the particles
together occupy the state \ket{0,0}{}.  The two detectors (the square boxes) are
initially brought together at event $I$ in order to prepare their variables as
specified in A\&A (Ref.\ 2, Eqs.\ 13, 18).  The detectors then separate for the
purpose of interacting with the particles at the space-like events 1 and 2.  After these interactions, the
detectors turn back to reunite with one another at event $M$ so their variables can be jointly compared. The
last step is not  discussed by A\&A, yet joint   comparison is essential.  Prior to this reunion the
separate detector variables are indefinite, which is why neither detector by itself can measure the spin
state of either particle.  However, the detectors are correlated by the initial preparation in such a way
that their \emph{combined} values are definite (Ref.\ 2, p.\ 362-3).  This combined definiteness applies before
event 1, and again  after event 2 in the  Lorentz frame of Fig.\ 1. It is this that allows an observer at
event
$M$ to make a definite measurement on the combined apparatus, confirming that the
particles remain in the state \ket{0,0}{}.  Aharonov and Albert call this a nonlocal \emph{nondemolition} experiment
because it measures a nonlocal state without destroying it.

	Of course, it is not necessary to being the detectors together if the information
can be communicated to event $M$ by some other means.  It is only necessary to insure
that such information is correctly combined \emph{prior} to an observation.   

\begin{figure}[t]
\centering
\includegraphics[scale=.90]{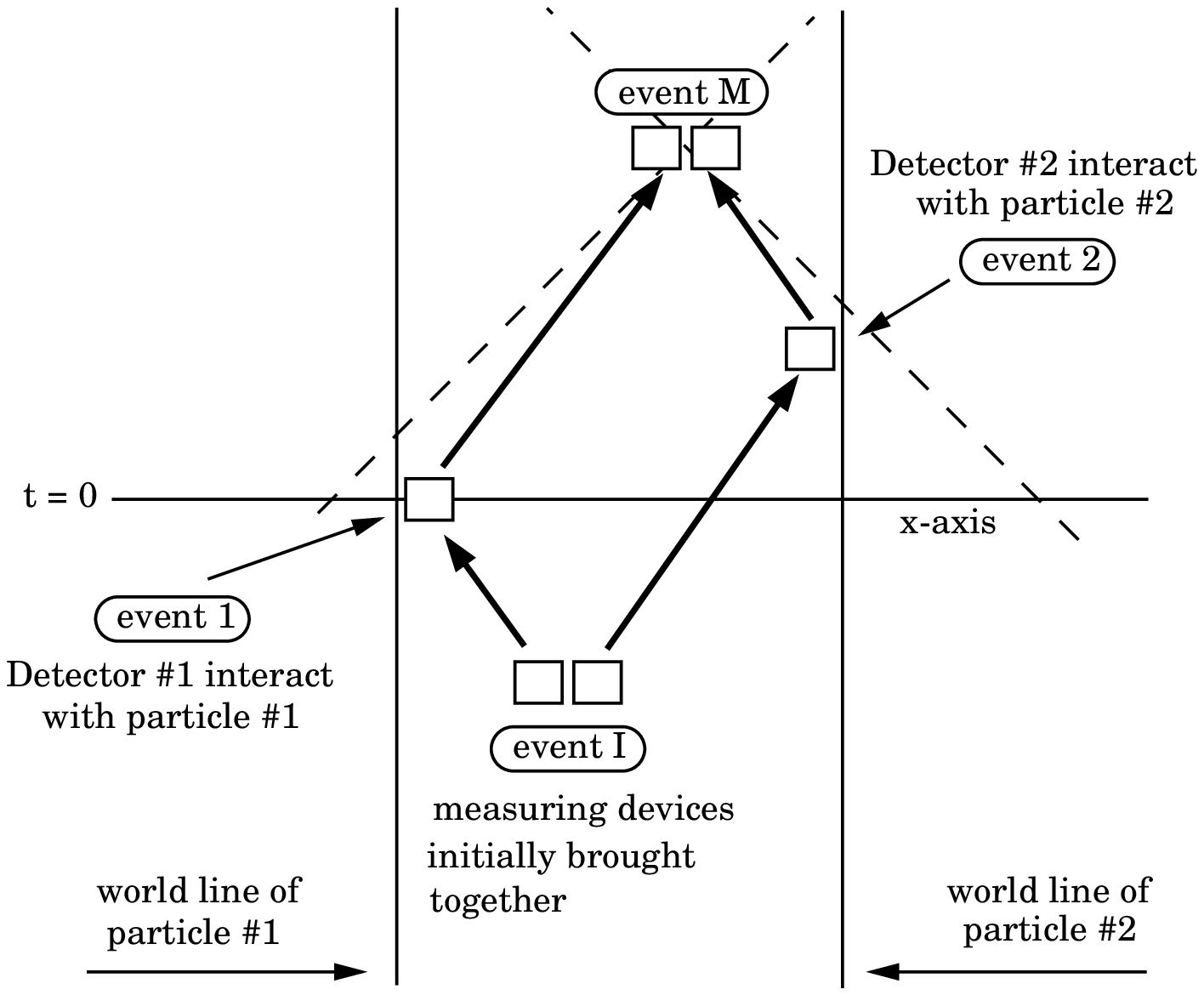}
\caption{}
\end{figure}

	The detector's combined variables will be correlated any time before event 1 and after event 2 as has
been said.  But A\&A show that this correlation is destroyed at any time \emph{between} events 1 and 2.  Between
these events they say,        ". . . the full state will not be separable into a state of the two-particle system
and a state of the measuring apparatus . . ." (Ref. 2, p.364).  The interaction at event 1 therefore disturbs the
system in such a way as to entangle the particle and detector states in Hilbert space, and this disturbance is
rectified by the interaction at event 2.  In a Lorentz frame in which events 1 and 2 are simultaneous, this
disturbance does not occur.  

\section{Sequential cycles}

	We now subject each detector to two interactions (see Fig.\ 2), where each detector moves along with, and
remains close to, the particle that it monitors.  The first detector on the left makes contact with
particle
\#1 (not shown) at events 1 and 3, and the second detector on the right makes contact with particle \#2 (not
shown) at events 2 and 4.  Both events on the right are assumed to have a space-like relationship to both
events on the left.  

 	These detectors have been initially prepared at an event such as $I$ in Fig.\ 1, but in this
case they are not reunited for the purpose of measurement.  Instead, a measurement is
affected by communicating information about the detector's variables to a common event like $M$ by
other means.  This information can be taken from any pair of events along the world lines of
the detectors, such as events $a$ and $v$, or $b$ and $u$.  We will say that the
\emph{detectors have been compared} at events $m$ and $n$ when information from these two
events has been brought together at the common event for comparison.  This comparison will not by itself
produce a state reduction.  We will say that \emph{the detectors have been observed} at events $m$
and $n$ when the comparison at those two events has been externally ``observed" at the common event. 
This observation will result in a reduction that is reflected back along the surface of the
backward time cone of the common event (through nonlocal correlations) to the detectors and the particles. 
These are reduced when their world lines penetrate the surface of the backward time cone of the
common event, as prescribed by Hellwig and Kraus.  

	As previously stated, when the detectors are compared at events $a$ and $u$ or $b$ and $v$, their combined
variables will be definite and they will register the fact that the particles occupy the state \ket{0,0}{}. If,
additionally, the detectors are \emph{observed} at events $a$ and $u$ or $b$ and $v$, then the particle state
\ket{0,0}{} will be empirically verified and will be identically reduced.

 We also learned in the previous section that if the detectors are compared at events $b$ and
$u$, their combined variables will be indefinite.  This reflects the fact that the system
is disrupted after event 1 and before event 2.  Therefore, an \emph{observation} of detectors at
events $b$ and $u$ will be accompanied by a state reduction that decouples the particles. 
Such an observation is irreversible.  It will disrupt the system in a way that event 2
cannot restore. 

\begin{figure}[t]
\centering
\includegraphics[scale=.90]{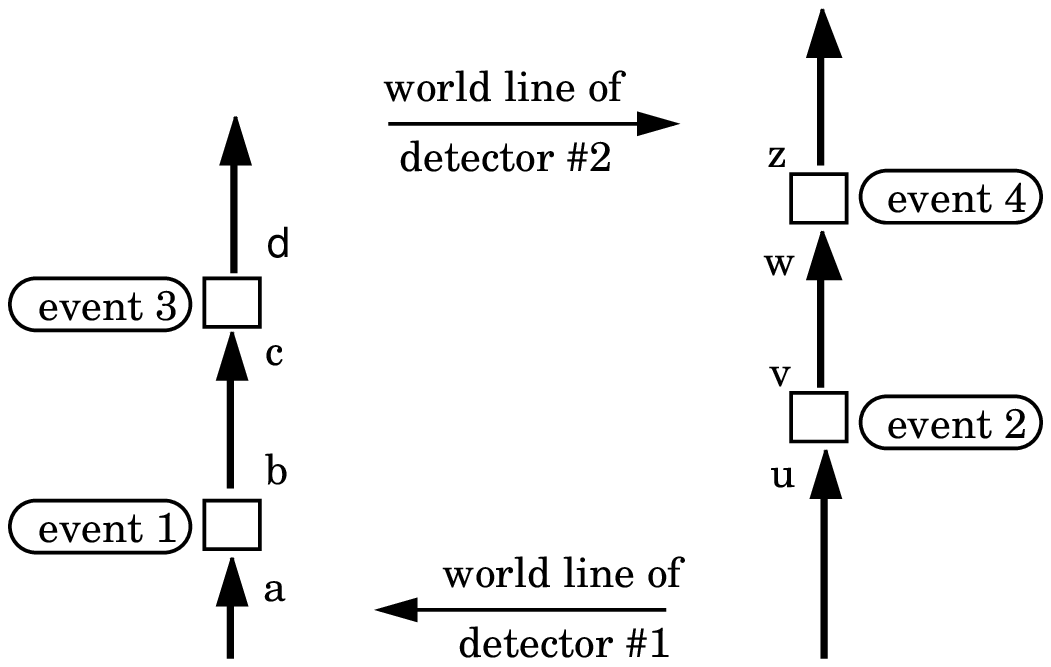}
\caption{}
\end{figure}

If the system survives events 1 and 2 without a disruptive observation, we will have
completed a nondemolition cycle.  Events 3 and 4, complete another nondemolition cycle. 
This cycle can be repeated many times so long as there is no permanent disruption of the
system due to an observation of the detectors at events such as $b$ and $u$, or $d$ and $w$. 

Since nothing happens to alter the states between event $w$ and $v$ (projecting backward in
time), the particle and detector states at events $d$ and $v$ are also non-separable in
Hilbert space.  This means that when the detectors are $compared$ at events $d$ and $v$,
their combined variables will be indefinite.  And if the detectors are $observed$ at events
$d$ and $v$, the resulting irreversible reduction will permanently disrupt the system,
destroying the state \ket{0,0}{} by decoupling the particles from one another.  This, we
say, happens because the pair of events 3 and 2 is \emph{demolitional}.   

	If Fig.\ 2 is Lorentz transformed in such a way as to make event 2 occur before event 1,
then by a similar reasoning the combined variables at events $b$ and $z$ will also be
indefinite.  Therefore, events 1 and 4 are demolitional as well.  It appears that
observations of either of the cross pairs 1 and 4, or 3 and 2, will irreversibly disrupt the
particle state, whereas parallel pair observations will not.  This conclusion is Lorentz
invariant.  

	All of the above assumes that the detectors are correctly prepared at event $I$.  With a different
initial preparation of the detectors, Aharonov and Albert show that events 3 and 2 can be made to be
nondemolitional, and event 1 and 2 would then be demolitional.  In this case, the change in initial
preparation is one that would cause detector \#1 to emerge from event 1 in such a way that a comparison of
the combined variables (after 1 and before 2) would be definite. 

	Aharonov and Albert draw a different conclusion from sequential experiments like those in Fig. 2. 
For them, the parallel event pairs 1-2, and 3-4 constitute complete nondemolition cycles that automatically 
\emph{include} state reduction, and this is supposedly achieved without having to introduce an external observer 
(Ref. 2, p. 365). They also say that the  cross event pairs 3-2, and 1-4, are complete cycles (again, including
state reduction without the benefit of an observer), but this time the result is demolitional.  As a result, we
have the odd situation that the four unobserved events in  Fig.\ 2 are said to leave the system in the original
state
\ket{0,0}{}, and at the same time, they are said to leave it irreversibly reduced in a disrupted state.  A\&A
accept this result at face value, allowing all such conflicting state reductions to be realized.  They formalize
the apparent contradiction by representing reduced states as surface functionals, thereby  allowing the different
reductions to apply along different temporal hypersurfaces. This can be done in such a way as to preserve Lorentz
invariance, which otherwise appears to be lost
\cite{AA2}. 

In this paper we say that  competing reductions are not equally realized.  A realized
reduction will be either demolitional or nondemolitional, and the one that actually occurs depends
on which events are witnessed by an outside observer.

\section{Two processes}

	John von Neumann's theory of measurement introduces a special non-Schr\"{o}dinger process,
which he calls Process I, that describes the collapse of the state function under
measurement.  It changes a pure quantum mechanical state into a classical mixture.  The
standard Schr\"{o}dinger evolution, which he calls Process II, can only change pure states into
pure states \cite{vonN}.  

	If the detectors in Fig. 2 are not observed, they will evolve under Schr\"{o}dinger (von
Neumann's Process II) causing their joint variables to vary back and forth between definite
and indefinite values.  This would produce no
paradox, no irreversible reduction, and no difficulty with the ordinary meaning of Lorentz invariance.  

We have seen in the previous section that detector information must be brought to a common location for
external examination.  It is here that we make the observation that initiates von Neumann's Process I, and
locates the vertex of a Hellwig-Kraus reduction at the observation site. Since we decide when and
where we are going to make an observation, we are the ones who decide whether the associated reduction will
preserve, or permanently destroy the particle state.  We can choose to observe  events 1 and 2, or 3 and 4, and
preserve the state \ket{0,0}{}; or we can choose to observe  events 3 and 2, or 1 and 4, and irreversibly destroy
the state.  

Choosing when to measure a quantum mechanical system is identical with deciding when to
impose additional boundary conditions.   It opens a closed system to further conditionals. 
It is difficult for many physicists to believe that there exists an independent process that
competes with Schr\"{o}dinger at such a fundamental level, but this is the implication of von
Neumann's theory of measurement.  Von Neumann showed that the boundary between a quantum
mechanical system and a measuring device is arbitrary.  However, there must always be a
boundary, and there is always something on the observer side of that boundary that cannot be
included in the system\footnote{This boundary has nothing to do with the
microscopic/macroscopic distinction.  According to von Neumann, a macroscopic measuring
device can always be included inside of a quantum mechanical system.}. If everything were includable, then
the wider system would be entirely Schr\"{o}dinger driven.  It  would then be unable to undergo a state
reduction\footnote{This statement precludes solutions of the type proposed by Ghiradi \emph{et al}
\cite{Ghi}}. Therefore, something that is intrinsically outside the system gives rise to
state reduction.  Von Neumann himself believed this ``something" to be related to the existence of
\emph{consciousness}, as have several others
\cite{Wig}\cite{Lon}\cite{Bea} including the author \cite{Mou}\cite{Mou2}.  But however
consciousness may or may not enter into the picture, reduction represents a boundary making
ability of the quantum mechanical observer that cannot be set aside. These boundaries are an
essential part of quantum epistemology inasmuch as a quantum mechanical universe without
definite boundaries is alien to human experience.  We humans are constrained to deal with
classical reality at the boundary of a quantum reality, and are therefore limited to a
theory that deals with  finite quantum mechanical systems that exist outside of ourselves. 
Heisenberg writes, ``. . . it is important to remember that in natural science we are not
interested in the universe as a whole, including ourselves, but we direct our attention to
some part of the universe and make that the object of our studies" \cite{Hei}.  There is
something about ourselves, as quantum mechanical observers, that places us firmly outside of
the Schr\"{o}dinger-driven world of quantum mechanics.  When this distinctive role of the
observer is taken into account, \emph{a la} von Neumann, the Aharonov and Albert experiment
is found to be fully consistent with the covariant reduction theory of Hellwig and Kraus.  

\part{}

\section{Causal loops}

	Some find a Hellwig-Kraus reduction disturbing in the way that it projects its influence
backward in time (relative to a given Lorentz frame), thereby raising the prospect of a
self contradictory causal loop (Ref.\ 3, p.1696).

	Imagine that the spin of the two particles in the state \ket{0,0}{} are observed by two different (and
separated) observers, where there is now no attempt to preserve the state as in Part I. 
Suppose the first particle is observed at event $A$ in Fig.\ 3, and is found to have a positive
spin.  Because of correlations, the second particle at event $B$ will then be
found to have a negative spin.       

 Neither particle can be found in the twice reduced region III, for
otherwise there would be a Lorentz frame in which one of the particles would be found in two
different places at once.  This means that the reduced state of event $A$ is confined to the
space-like region around event $A$ that overlaps the backward time cone of event $B$ (labeled region
II in Fig.\ 3).  Similarly, the reduced state of event $B$ is confined to region I in Fig.\
3.  We ignore what happens in the forward time cones of events $A$ and $B$.  

\begin{figure}[t]
\centering
\includegraphics[scale=.90]{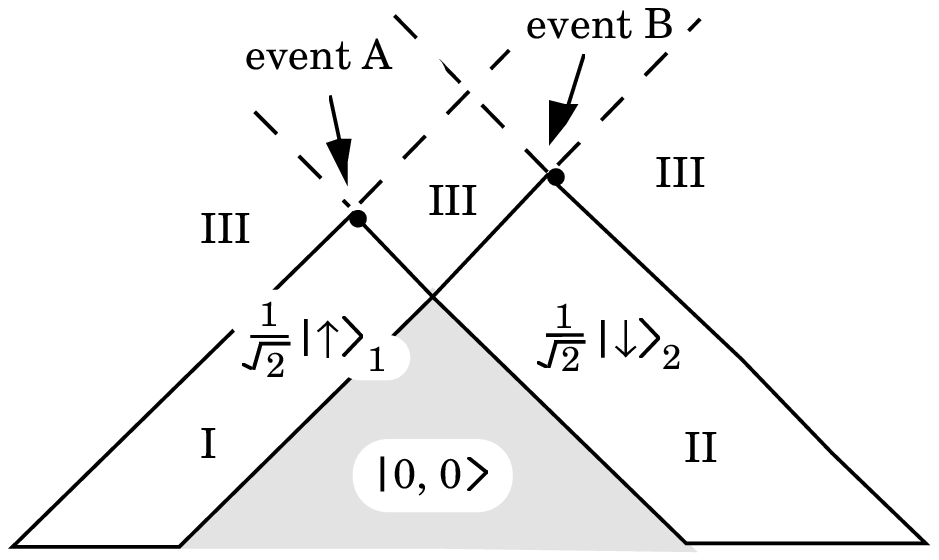}
\caption{}
\end{figure}

We can find the unrenormalized reduced state of event $A$ by operating on \ket{0,0}{} in Eq.\ 1
with \bra{\uparrow}{1}, the measured spin of the first particle at $A$.

\begin{equation}
\bra{\uparrow}{1}\ket{0,0}{}\ =\ \frac{1}
{\sqrt{2}}\ket{\downarrow}{2}
\end{equation}
This reduced single particle ket occupies the immediate backward time cone of event $B$ as
shown in region II of Fig.\ 3.  It tells us that the observation at event $A$ leaves only one
spin possibility at event $B$.\footnote{von Neumann's non-relativistic Process I for this particular
\emph{re}normalized reduction gives the state \ket{\uparrow}{1}\ket{\downarrow}{2} that exists any time after
event $A$ and before event $B$ in the Lorentz frame of Fig. 3.  This is compatible with the relativistic H-K
result if we let \ket{\uparrow}{1} be limited to the future time cone of event $A$, and let \ket{\downarrow}{2} be
limited to the backward time cone of event $B$.}  When the second particle is subsequently observed at
$B$, the result is

\begin{equation}
\bra{\uparrow}{1}\bra{\downarrow}{2}\ket{0,0}{}\ =\ \frac{1}
{\sqrt{2}}      
\end{equation}
whose square magnitude is 0.5, the joint probability of finding the first particle with spin
up, and the second particle with spin down.  The reduced state of event $B$ (shown in region
I) can be found in a similar way.  

	What may be disturbing about this theory is that the reduced state in the immediate
backward time cone of event $B$ in Fig.\ 3 is a function of what happens at event $A$.  If the
observer at $A$ decides to look at his apparatus, that decision clearly influences the form of
the reduced state leading into event $B$.  One might therefore suppose that observer A could send a
superluminal message to observer B by deciding not to look.  But that will not work.  When B records
a negative spin, he does not know if that happens because he is looking at the original
state \ket{0,0}{} and just happens to measure `spin down', or because he is looking at the reduced
state of event $A$ which makes `spin down' the only possibility at event $B$.  Therefore, A's decision
to observe, or not, cannot send a decipherable message to observer B.  

	Although observer A can choose to make an observation, he cannot choose the outcome; and so, he cannot choose B's
outcome either.  He can no more sent a message to B than he can send one to himself by this means.	The
fact that B's outcome  depends on A's outcome is called \emph{outcome dependence}, and the fact that B's
result is independent of A's decision to make an observation is called \emph{parameter independence}
(Ref.\ 4, Sect.\ 8.2).  

An ensemble of experiments of this kind is also unable to communicate superluminally.  The
observer at event $A$ can only record the probability of finding `spin up', and he
will get 0.5 no matter what the observer at event $B$ does or experiences.  In order to confirm that
the joint probability is 0.5 for the ($\uparrow,\downarrow$) spin combination, it will be
necessary for observers A and B to get together (at some later time) to compare notes - to verify
correlations.  Superluminal communication is then no longer a consideration.  

\section{Non-local Boundaries}

	There is still another objection that is raised in connection with the reduction scheme in
Fig.\ 3.  The two observations at events $A$ and $B$ in that figure are a pair of boundary conditions that
are placed on the original state \ket{0,0}{}.  However, neither one is anywhere near the boundary of that
function.  They are both some distance away from the shaded area in Fig.\ 3 over which they
have joint jurisdiction.  Cohen \& Hiley comment on this consequence of the Hellwig-Kraus
theory saying, ``. . . the two-particle wave function is reduced before either particle has
been subjected to a measurement, which seems absurd" (Ref.\ 3, p.\ 1695).   The author
disagrees.  

	We know that the correlations found in this two particle spin system act nonlocally through
distance.  There is nothing absurd about that, or at least, there is nothing new about that
since the discovery of Bell's inequalities \cite{Bell}.  Certainly these correlations can be
nonlocal through time as well as distance in any Lorentz frame - so long as they do not extend into the
backward time cone of a given measurement.  There is therefore no
reason why the \emph{joint} events A and B in Fig.\ 3 should not be the terminal boundary conditions
of the shaded area in that figure, as well as the initial conditions of the reductions in regions I and
II.

\end{document}